\documentclass[12pt]{iopart}

\expandafter\let\csname equation*\endcsname\relax
\expandafter\let\csname endequation*\endcsname\relax

\usepackage{dcolumn}
\usepackage{bm}
\usepackage[colorlinks=true]{hyperref}
\usepackage[]{amsmath}
\usepackage{amssymb}
\usepackage{pifont}
\usepackage{revsymb}

\usepackage{graphics}
\usepackage{graphicx}
\graphicspath{{../figures/pdfs/}}

\usepackage{soul}
\usepackage{multirow}
\usepackage[dvipsnames]{xcolor}

\usepackage{color}
\usepackage{tablefootnote}

\usepackage{cancel}

\newcommand{\BD}{{\mbox{\tiny BD}}}

\newcommand{\WD}{{\mbox{\tiny WD}}}

\newcommand{\PSR}{{\mbox{\tiny PSR}}}
\newcommand{\PSRWD}{{\mbox{\tiny PSR-WD}}}
\newcommand{\WDWD}{{\mbox{\tiny WD-WD}}}
\newcommand{\NSNS}{{\mbox{\tiny NS-NS}}}

\newcommand{\boldhat}[1]{\hat{\mathbf{#1}}}

\begin{document}

\title{Probing Massive Scalar Fields from a Pulsar in a Stellar Triple System}

\author{Brian C. Seymour}
\address{Department of Physics, University of Virginia, Charlottesville, Virginia 22904, USA}

\author{Kent Yagi}
\address{Department of Physics, University of Virginia, Charlottesville, Virginia 22904, USA}

\date{\today}

\begin{abstract} 


Pulsar timing observations precisely test general relativity. Recently, the hierarchical triple system PSR J0337+1715 has placed new constraints on the  existence of a fifth force from violation in the strong equivalence principle. Many alternative gravity theories exist with massive (pseudo-)scalar fields to explain a variety of phenomena from the accelerating expansion of the universe at large scales to the QCD strong CP problem at small scales. We here develop a generic formalism for the fifth force effect in theories involving massive scalar fields arising from e.g. string theory. With PSR J0337 measurements, we find the strongest bound on the simplest theory with a massive scalar field beyond general relativity and derive new constraints in other theories with axions, dark matter mediators, and higher-curvature corrections. These results show that the triple system J0337 provides a stringent test for massive scalar fields.

\end{abstract}
\maketitle

\section{Introduction} 
In the late 16th century, Galileo allegedly showed that all masses feel gravity equally in his famous Leaning Tower of Pisa experiment. This suggested that all test masses experience free fall when gravity is the only physical force involved. Such an experiment also hinted at a central pillar of gravitational theory, the equivalence principle, that Einstein would use centuries later. The \emph{strong} equivalence principle (SEP) is at the core of general relativity (GR): all test masses -- including self-gravitating ones like stars and black holes -- feel universal gravitation. 

Currently, GR has been extensively verified with various experiments and observations~\cite{Will:2014kxa,Berti:2015itd}. Solar system tests have thoroughly constrained the weak field regime of gravity~\cite{Will:2014kxa}. Binary pulsar measurements have also made precision tests of the non-dynamical, strong field regime through pulsar timing~\cite{Wex:2014nva,Stairs:2003eg}. Cosmological observations can probe the large-scale nature of gravity~\cite{Koyama:2015vza}. Additionally, gravitational wave observations have probed the dynamical/strong field regime of GR~\cite{TheLIGOScientific:2016src,Yunes:2016jcc,Abbott:2018lct,LIGOScientific:2019fpa,Nair:2019iur,Yamada:2019zrb}. However, 
we can only know the validity of GR by continually testing it at ever higher precision and energy scales. Moreover, since GR cannot be made into a quantum theory, it is only an effective field theory description of nature, and it must break down at some scale where quantum effects become relevant \cite{Will:2014kxa,Berti:2015itd}.

In this paper, we focus on testing theories with extra dynamical scalar fields which arise in many different contexts. For example, string theory predicts a plethora of scalar fields in the form of dilatons and moduli. Moreover, its low-energy effective theory leads to scalar-tensor theories~\cite{Fujii:2003pa}. Scalar fields can also source inflation and cosmic acceleration~\cite{Clifton:2011jh,Jain:2010ka}. Furthermore, string theory predicts the existence of \emph{massive} (pseudo-)scalar fields called axions. These axions are especially interesting because they are also candidates for cold dark matter~\cite{Duffy:2009ig}. Massive scalar fields also arise in certain modified theories of gravity within the context of scalar-tensor theories~\cite{Alsing:2011er,Berti:2012bp} and as dark matter mediators~\cite{Alexander:2018qzg}. 
However, these additional scalar fields come with an important cost: the extra degrees of freedom, in general, give rise to the \emph{fifth} force (on top of the four known forces in physics) between two self-gravitating objects. This force depends on the internal structure of each body and may violate SEP~\cite{Will:2014kxa}. We also note that there have been mathematically rigorous approaches to studying Einsteinian gravity coupled to extra scalar fields \cite{LeFloch:2015ppi,Wyatt:2017tow,Branding:2018xsr}, but we will focus in this paper on a phenomenological approach to modified gravity.

Previous experiments have placed stringent constraints on SEP violation within the solar system. For example, Lunar Laser Ranging (LLR) allows us to measure the difference in the gravitational acceleration of the Earth and the Moon towards the Sun
~\cite{Merkowitz:2010kka,Hofmann_2018,2010A&A...522L...5H} (the so-called Nordtvedt effect~\cite{Nordtvedt:1968qr}), which should vanish if the gravity acts onto objects \emph{universally} as in GR. Furthermore, NASA MESSENGER has measured the orbit of Mercury very precisely and bounded SEP violation even more stringently than LLR experiments~\cite{2018NatCo...9..289G}.

\begin{figure}[thb]
    \centering
    \includegraphics[width=0.6\linewidth]{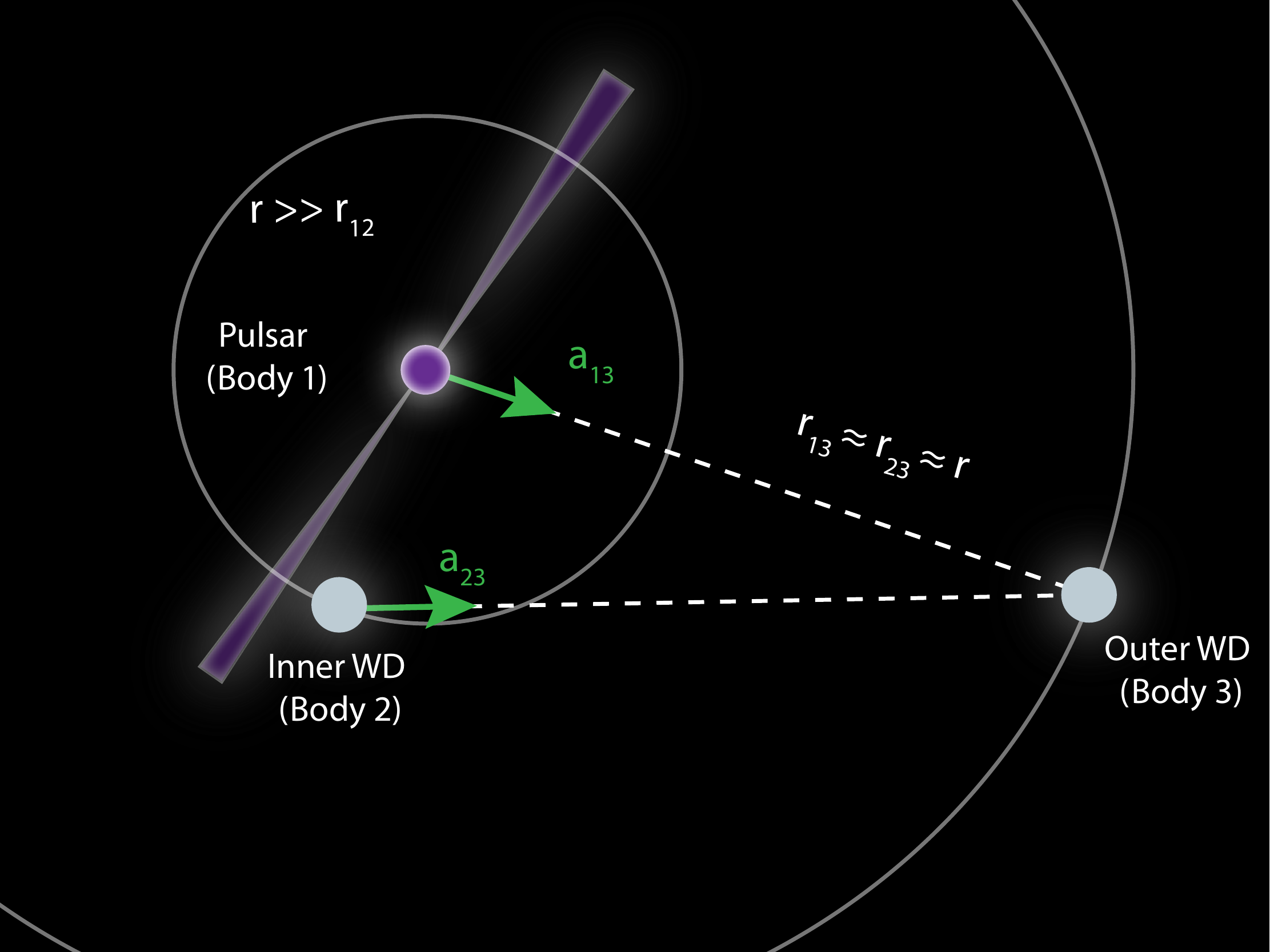}
    \caption{
    Schematic picture of the triple system PSR J0337+1715~\cite{Archibald:2018oxs}. The outer WD orbits the inner binary consisting of a pulsar and an inner WD. The violation of SEP is characterized by the difference in the acceleration of the pulsar ($\bm a_{13}$) and inner WD ($\bm a_{23}$) towards the outer WD ($\bm a_{13} = \bm a_{23}$ in GR). Its fractional difference has been constrained to be less than $2.6\times 10^{-6}$~\cite{Archibald:2018oxs}.
    }
    \label{fig:diagram}
\end{figure}

Binary pulsar observations can probe SEP violation even more accurately. In particular, one can now carry out a similar test to LLR thanks to the recent discovery of the pulsar in a stellar triple system PSR J0337+1715~\cite{Ransom:2014xla,Shao:2016ubu} (``PSR J0337'' hereafter). This system consists of an inner millisecond pulsar-white dwarf (WD) binary and a second WD in an outer orbit (Fig.~\ref{fig:diagram}). The SEP violation was constrained by comparing the acceleration due to the outer WD onto the pulsar and inner WD (shown as green arrows $\bm{a}_{13}$ and $\bm{a}_{23}$ in Fig.~\ref{fig:diagram}) \cite{Archibald:2018oxs}. The new measurement of the SEP violation improves on the previous measurements from NASA Messenger~\cite{2018NatCo...9..289G} and LLR~\cite{Hofmann_2018,2010A&A...522L...5H} a factor of 5 and 7 respectively. Unlike the previous solar system tests of SEP violations that can only probe weak-field effects, PSR J0337 probes strong-field effects since both neutron stars and WDs are strongly self-gravitating objects. Indeed, new bounds on the SEP violation from PSR J0337 have been used to place the most stringent bounds on the massless scalar-tensor theories~\cite{Archibald:2018oxs}. 

In this paper, we apply the SEP violation bound from PSR J0337 to theories with massive scalar fields. We first derive a generic formulation that describes the Nordtvedt effect in such theories. We then use this generic formulation to derive bounds on example theories.

Regarding theories beyond GR, we consider (i) Horndeski theory~\cite{Horndeski:1974wa,Kobayashi:2019hrl}, (ii) massive Brans-Dicke theory, and (iii) metric $f(R)$ gravity. Horndeski theory is the most general scalar-tensor theory of gravity that contains at most 2nd order derivatives in its field equations. Massive Brans-Dicke theory~\cite{Alsing:2011er,Berti:2012bp} is an extension to the well-known Brans-Dicke theory~\cite{Brans:1961sx} endowing the scalar field with a non-vanishing mass. Metric $f(R)$ gravity is a higher-curvature corrected theory where Ricci scalar $R$ in the Einstein-Hilbert action for GR is replaced by a function of the Ricci scalar $f(R)$. This last theory can be mapped to massive Brans-Dicke theory~\cite{Sagunski:2017nzb}.

We also consider theories involving massive scalar fields within GR. Of course, these theories do not violate SEP, but they can give rise to a fifth force due to a new interaction caused by the massive scalar fields. One example that we study in this paper is the axion. Axions were originally introduced to solve the strong CP problem~\cite{Peccei:1977hh} in QCD\footnote{The strong CP problem asks why QCD preserves CP symmetry. Experiments seem to measure CP symmetry in QCD (e.g. by measurements of neutron electric dipole moment) while the mathematical description of it does not generically have CP symmetry. This is curious because QCD must have a finely tuned parameter to have CP symmetry.}. As already mentioned, they also arise from string theory, creating many interesting phenomena in a string axiverse~\cite{Arvanitaki:2009fg}. Both QCD and string axions are candidates for cold dark matter. The second example theory of a massive scalar field within GR is the gravitationally-bound dark matter inside stars~\cite{Goldman:1989nd,Kouvaris:2010vv,Alexander:2018qzg}. Regardless of the dark matter model, a Yukawa modification can arise for the gravitational potential of a neutron star if a fifth force is caused by an extra light force mediator for dark matter~\cite{Alexander:2018qzg}. 

Below, we use the geometric units $c=G=1$ unless otherwise stated.

\section{Formulation}

Let us now present a new framework to capture the existence of a fifth force (that may lead to a violation of SEP) due to massive scalar fields in a generic way. We will see that Yukawa potentials arising generically from massive scalar fields result in emergence of SEP violation in a three body problem. SEP violation has been computed in specific theories previously (for Massive Brans-Dicke \cite{Alsing:2011er} and Horndeski \cite{Hou:2017cjy}), but we will derive its expression in other scalar fifth force theories ($f(R)$ gravity, axions, and an example DM theory). Furthermore, we have generalized this with a novel parameterized expression for SEP violation with massive scalar fields.

Let us first see the modification of Newtonian acceleration with the presence of a massive scalar field. The field equations for a scalar field will be of the form, 
\begin{equation}
    \left( \Box - m_s^2 \right)\phi = S \, ,
\end{equation}
where $m_s$ is scalar field mass, $\Box$ is the d'Alembertian operator, $\phi$ is the scalar field, and $S$ is the sources. From this, we see that a scalar field will generically create a Yukawa potential due to the Green's function of $\Box - m_s^2$.
A scalar field creates a Yukawa potential between objects with (inertial) masses $m_A$ and \emph{dimensionless} scalar charges $q_i$ ($i=1, \,2$) as 
\begin{equation}\label{eq:YukawaPotential}
    V_\phi = -B m_1m_2 \frac{ q_1 q_2}{r}e^{- r/\lambdabar} \, ,
\end{equation}
where $\lambdabar \equiv \hbar/(m_s c)$ is the reduced Compton wavelength of the massive scalar field, and $B$ is a theory-dependent dimensionless coupling constant. 
This, together with the Newtonian potential ($V_N \equiv -m_1 m_2 /r$) \footnote{In this work, we renormalize the gravitational constant for the potential with no scalar charges. Note that we are setting this renormalized gravitational constant to be 1, not the bare gravitational constant.} can be unified into a single potential as 
\begin{align}
    V_{ij} \equiv -\frac{m_i m_j}{r_{ij}}\mathcal{G}_{ij} \, , \quad
    \mathcal{G}_{ij} \equiv 1 + B q_i  q_j e^{- r_{ij}/\lambdabar} \, , \label{eq:generalpotential}
\end{align}
where we define $\bm r_{ij} \equiv \bm r_i - \bm r_j $ and $r_{ij} \equiv \mid \bm r_i - \bm r_j \mid$ is the separation between objects $i$ and $j$. The acceleration of the $i$th object is found by taking the gradient of Eq.~\eqref{eq:generalpotential}, which yields 
\begin{equation}\label{eq:genericacc}
    \bm a_i = - \frac{1}{m_i} \nabla_i \sum_{i\ne j} V_{ij}= -\sum_{j\neq i} \frac{m_j}{r_{ij}^2}\left[\mathcal{G}_{ij} +B   q_i q_j \frac{r_{ij}}{\lambdabar}e^{- r_{ij}/\lambdabar}\right] \boldhat r_{ij} \, ,
\end{equation} 
where we define the unit vector $\boldhat r_{ij} \equiv \bm r_{ij} / r_{ij}$. Note that the acceleration of body $i$ depends on the scalar charge $q_i$. This breaks the equivalence principle.

Now, we will specialize the $n$ body acceleration equation to a hierarchical triple system. In order to obtain the fifth force parameter from our expression for acceleration, we must consider the difference in acceleration $\bm a_{12} \equiv \bm a_1 -\bm a_2$ that bodies 1 and 2 experience from a third~\cite{Alsing:2011er,Will:2018ont}. For hierarchical triple systems like the Earth-Moon-Sun system and PSR J0337 where the outer body is much further away from the other two, one can impose $r_{12}\ll r_{13}$ and $r_{13}\sim r_{23} \sim r$, where $r$ is distance from the inner binary's center of mass to body 3\footnote{To see a similar analysis to higher perturbation order, it can be found in section 8.1 of Ref.~\cite{Will:1993hxu}. It is noteworthy that tidal interactions enter at higher orders.}. Assuming further $r_{12} /\lambdabar \approx 0$,\footnote{We have checked that all our bounds are within this approximation regime.} the expressions for $\bm a_1$ and $\bm a_2 $ are
\begin{align}
    \bm a_1 &= -  \frac{\mathcal{G}_{12} m_2}{r_{12}^2}  \boldhat r_{12} - \left( \frac{ m_3}{r^2}\right)\left[ \mathcal{G}_{13} + B q_1 q_3 \frac{r}{\lambdabar} e^{- \frac{r}{\lambdabar}} \right] \boldhat r \, , \\
    \bm a_2 &= +  \frac{\mathcal{G}_{12} m_1}{r_{12}^2}  \boldhat r_{12} - \left( \frac{ m_3}{r^2}\right)\left[ \mathcal{G}_{23} + B q_2 q_3 \frac{r}{\lambdabar} e^{- \frac{r}{\lambdabar}} \right] \boldhat r \, .
\end{align}
Now, we find the expression for $\bm a_{12}$
\begin{equation}\label{eq:diffacceleration}
    \bm a_{12} = - (1+q_1 q_2 B) \frac{m_1+m_2}{r_{12}^2} \boldhat r_{12}  - \frac{m_3}{r^2}\Delta \,\boldhat r  \, .
\end{equation} 
The first term is the rescaled Newtonian acceleration and the second is due to the fifth force. The fifth force creates a relative acceleration between the inner bodies towards the third body. Thus, a fifth force would create SEP violation in this astrophysical system.

Let us now examine the fifth force parameter $\Delta$. For the pulsar in a triple system like PSR J0337, the fifth force parameter due to massive scalar fields can generically be expressed as 
\begin{align}\label{eq:sep}
    \Delta &= \left(1 + \frac{r}{\lambdabar} \right)(\mathcal{G}_{13}-\mathcal{G}_{23}) \nonumber \\
    &=  B \left(1+ \frac{r}{\lambdabar}\right) (q_1 - q_2)q_3 e^{- r/\lambdabar}\,.
\end{align}
This is the generic expression for the fifth force parameter involving a massive scalar field. When fifth force is absent, GR guarentees that there is no SEP violation so we will have $\Delta = 0$. We test the presence of fifth force due to massive scalar fields by comparing how closely $\Delta$ is constrained to zero (with an experimental uncertainty of $\sigma_\Delta$).

Next, let us test the presence of massive scalar fields with the SEP-violation constraints from PSR J0337. First, we derive the fifth force parameter $\Delta$ in example theories and list it in Table~\ref{tab:Theories}. Below, we look at each theory in detail and use the expression $|\Delta| < 2 \sigma_\Delta$ to test a fifth force at 2-$\sigma$ level. We consider constraints with the measurement precision of $\sigma_\Delta = 2.6 \times 10^{-6}$ from PSR J0337~\cite{Ransom:2014xla,Archibald:2018oxs}. 

\begin{table}[thb]
    \centering
    \setlength\tabcolsep{1.5pt}
    \footnotesize
    \begin{tabular}{|c|c|c|c|c|c|}
    \hline
         Theory & Th.~Params  & $B$ & $q_i$ & $\Delta$ for J0337 & Refs. \\ \hline \hline
        Massive Brans-Dicke & ($\omega_\BD$, $m_s$)  & $\frac{1}{3+2\omega_\BD}$ & $1-2s_i $ & $-\frac{2}{3+2\omega_\BD}(1+ \frac{r}{\lambdabar}) s_1 e^{- r/\lambdabar} $ & \cite{Berti:2012bp}\\ \hline
        Quadratic $f(R)$ & $m_s=\sqrt{\frac{1}{6 \bar a_2}}$ & $ \frac{1}{3}$ & $1-2s_i$ & $-\frac{2}{3}(1+\frac{r}{\lambdabar})s_1 e^{- r/\lambdabar}$ & \cite{Starobinsky:1980te,Sagunski:2017nzb} \\ \hline 
        Horndeski & --  & $\frac{1}{G_{4(0,0)}(1+G_{4(0,0)}) \zeta}$ & $G_{4(1,0)} -\frac{2 s_i}{\phi_0}G_{4(0,0)}$ &$\frac{(q_1 - q_2)q_3}{G_{4(0,0)}(1+G_{4(0,0)}) \zeta}(1+\frac{r}{\lambdabar})e^{- r/\lambdabar}$  & \cite{Horndeski:1974wa,Hou:2017cjy}
         \\ \hline \hline
         Axion & ($f_a$, $m_s$) & $\frac{1}{4 \pi}$ & $ -\frac{8 \pi  f_a}{ \sqrt{\hbar}\ln{\left(1-\frac{2m_i }{R_i}\right)}} $& $-\frac{4 \pi  f_a^2}{\hbar} (1+\frac{r}{\lambdabar}) \frac{R_2R_3}{m_2m_3}e^{- r/\lambdabar} $& \cite{Hook:2017psm,Huang:2018pbu}\\ \hline
         Dark Matter & ($\alpha$, $m_s$)  & $1$ & -- & $ (1+ \frac{r}{\lambdabar}) \alpha_{\PSRWD} e^{- r/\lambdabar}$ & \cite{Alexander:2018qzg}\\ \hline
    \end{tabular}
    \caption{
    Mapping between fifth force parameters in Eq.~\eqref{eq:sep} and theoretical parameters in example theories. The first class represents theories beyond GR while the second class shows theories with scalar fields within GR. The first and second columns list example theories and their theoretical parameters. The third and fourth columns show the mapping for $B$ and scalar charge $q$. The last column shows $\Delta$ specific to J0337. Horndeski theory contains arbitrary functions instead of theoretical parameters and how the generic expression for $\Delta$ reduces to a simpler expression for J0337 depends on such functions. Note that $f(R)$  corresponds to MBD with $\omega_\BD = 0$. In our listing for axion charge, we assume that the star is at the critical density to source the axion field (below this cutoff the field is not sourced).
    }
    \label{tab:Theories}
\end{table}

\section{Results}
\subsection{Massive Brans-Dicke}
Massive Brans-Dicke theory is constructed by adding a potential $M(\phi)$ to the massless Brans-Dicke action. This forces the scalar field to acquire a scalar mass (squared)
\begin{equation}
    m_s^2 = -\frac{\phi_0}{3+2 \omega_\BD} M''(\phi_0) \, ,
\end{equation}
where $\omega_\BD$ is the Brans-Dicke parameter, and its inverse roughly specifies the coupling between the scalar field and matter. One recovers GR in the limit $\omega_\BD \to \infty$. The background value of the scalar field is $\phi_0=(4+2\omega_\BD)/(3+2\omega_\BD)$~\cite{Berti:2012bp}. The scalar field changes the effective gravitational constant to be $G = \phi_0/\phi$~\cite{Berti:2012bp}. Due to the dependence of the gravitational constant on the scalar field, we can define scalar charges $q_i = 1-2s_i$  of body $i$ for massive Brans-Dicke theory, where the sensitivity $s_i$ is defined as $s_i \equiv - \left.\partial(\ln \, m_i)/\partial(\ln \, G)\right|_{\phi_0}$. 
Following~\cite{Alsing:2011er}, we choose $s_\WD \ll s_\PSR = 0.2$. 
The Nordtvedt parameter $\eta_N$ characterizes SEP violation and is derived in~\cite{Alsing:2011er} for massive Brans-Dicke theory. 
One can easily find the fifth force parameter $\Delta$ by the relation $\Delta = \eta_N (s_2 - s_1)$.

\begin{figure}[thb]
    \begin{center}
        \includegraphics[width=0.6\linewidth]{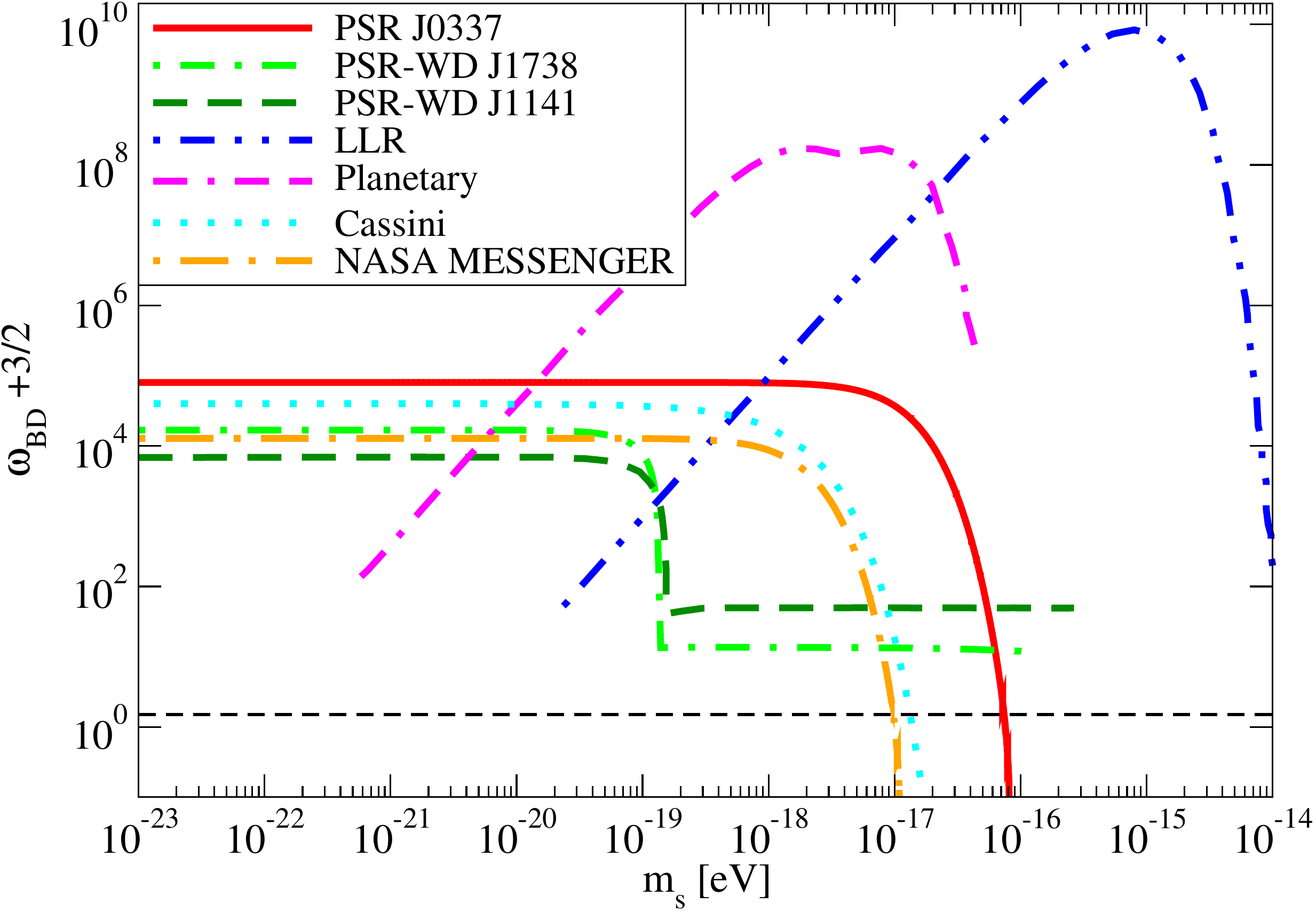}
        \caption{\label{fig:MBDOmegaBound} 
        The lower bound on the Brans-Dicke parameter $\omega_\BD$ as a function of the scalar field mass. We show the pulsar timing measurements of the SEP violation with PSR J0337 (red solid) and of the orbital decay rate in two pulsar/WD binaries~\cite{Dyadina:2018ryl,Alsing:2011er} (green dotted-dashed and green dashed). We also show the solar system measurements of Shapiro time delay by Cassini~\cite{Alsing:2011er} (cyan dotted) and of SEP violation by MESSENGER~\cite{2018NatCo...9..289G} (orange dotted-dashed). We include the inverse square law constraints from LLR (blue dotted-dashed) and planetary (magenta dotted-dashed) \cite{Bergebergberge:2017gxr,Adelberger:2009zz}. The horizontal line (black dashed) corresponds to the $f(R)$ gravity with $\omega_\BD = 0$. Larger values on $\omega_\BD$ corresponds to stronger bounds. Observe the new constraints derived here (PSR J0337, LLR, and planetary bounds) place the most stringent bound in massive Brans-Dicke theory. Note that the theoretical lower bound is $\omega_\BD > -3/2$, which is why we plot $\omega_\BD+3/2$ following the example of \cite{Alsing:2011er}. 
        }
    \end{center}
\end{figure}

Let us now examine the constraints arising from the pulsar triple system. Using the constraints on SEP violation with the triple system in conjunction with our expression for $\Delta$ in Table~\ref{tab:Theories}, we can construct the region in the parameter space $(\omega_\BD,m_s)$ ruled out by observations. Figure~\ref{fig:MBDOmegaBound} presents the lower bound on $\omega_\BD +3/2$ as a function of the scalar mass $m_s$. For example, the red solid curve is obtained as a contour corresponding to $\Delta = 2 \sigma_\Delta$, where the expression for $\Delta$ for massive Brans-Dicke theory can be found in Table~\ref{tab:Theories} and $\sigma_\Delta = 2.6 \times 10^{-6}$ for PSR J0337 as already mentioned. The new result from PSR J0337 now provides the most stringent bound on massive Brans-Dicke theory when the scalar field mass is sufficiently small and improves significantly on the previous strongest bound from the Cassini mission via Shapiro time delay. When the mass is relatively large, most stringent bounds come from LLR and planetary measurements, which are obtained here for the first time.

\subsection{Metric $f(R)$}
Next, let us consider metric $f(R)$ gravity. In particular, we consider a simple quadratic form $f(R) = R+\bar a_2 R^2$\footnote{Normally, the literature refers to $\bar a_2$ as $a_2$, but we add the bar to distinguish from acceleration $\bm a_2$.} which is motivated from e.g. inflation~\cite{Starobinsky:1980te}. Introducing an effective scalar field $\phi = df/dR$, one can show that this theory is equivalent to Brans-Dicke theory with $\omega_\BD = 0$ and a scalar mass $m_s = \sqrt{1/6\bar a_2}$. 
The black dashed line in Fig.~\ref{fig:MBDOmegaBound} corresponds to the parameter space of $f(R)$ gravity. One can thus see that all scalar mass values other than $m_s\geq8.2\times10^{-17} \text{eV}$ are ruled out by PSR J0337 bounds. This corresponds to a constraint of $\bar a_2 \leq 9.6\times 10^{17} \text{m}^2$. 

Let us compare this with other existing bounds. The earth-based experiments can probe the Yukawa correction to the inverse square law at small length scales. For example, the E\"{o}t-Wash experiment placed a considerably stronger bound with $\bar a_2 <10^{-10} \text{m}^2$~\cite{Kapner:2006si}. On the other hand, the double-pulsar binary PSR J0737-3039 constrains $\bar a_2 \lesssim 2.3\times 10^{15} \text{m}^2$ and Gravity Probe B yields $\bar a_2 \lesssim 5\times 10^{11} \text{m}^2$~\cite{Naf:2010zy}, while GW170817 gives $\bar a_2 \lesssim 4\times 10^{6} \text{m}^2$~\cite{Sagunski:2017nzb}. Thus, the new bound from PSR J0337 is weaker than any of the above and is not well suited for probing this particular $f(R)$ theory of gravity.

\subsection{Horndeski gravity}
Let us now briefly summarize Horndeski gravity and find its generic fifth force parameter. Horndeski gravity is the most general scalar-tensor theory of gravity with up to 2nd order derivatives in the field equations~\cite{Horndeski:1974wa}. The theory contains four arbitrary functions $G_i(\phi,X)$ for $i=(2,3,4,5)$ with $\phi$ representing the scalar field, $X \equiv -1/2 \, \phi_{; \mu} \phi^{; \mu}$ and  $ \phi_{; \mu} \equiv \nabla_\mu \phi$, and the action is given by 
\begin{equation}
S=\frac{1}{16 \pi} \sum_{i=2}^{5} \int d^{4} x \sqrt{-g} \mathcal L_{i}+S_m[ g_{\mu \nu}, \psi_{m}] \, ,
\end{equation}
where $g_{\mu\nu}$ is the metric while $S_m$ is the action for matter field $\psi$~\cite{Kobayashi:2011nu}. $\mathcal L_i$ is defined to be

\begin{align} 
\mathcal L_{2}=& G_{2}(\phi, X)\,, \\ 
\mathcal L_{3}=&-G_{3}(\phi, X) \square \phi \,,\\ 
\mathcal L_{4}=& G_{4}(\phi, X) R+G_{4 X}\left[(\square \phi)^{2}-\left(\phi_{;\mu\nu}\right)^{2}\right] \,,\\ 
\mathcal L_{5}=& G_{5}(\phi, X) G_{\mu \nu} \phi^{;\mu \nu}-\frac{G_{5 X}}{6}\left[(\square \phi)^{3}+2\left(\phi_{;\mu \nu}\right)^{3}-3\left(\phi_{;\mu \nu}\right)^{2} \square \phi \right]\,,
\end{align}
with $G_{i X} = \frac{\partial G_i}{\partial X}$ and $\square \equiv g^{\mu \nu} \nabla_\mu \nabla_\nu$.

We now derive the expression for the fifth force parameter $\Delta$ in Horndeski gravity. From the expression for the relative acceleration of two bodies in Eq.~(44)\footnote{Our expression in Eq.~\eqref{eq:SEPEq} differs from Eq.~(44) of Ref.~\cite{Hou:2017cjy} because we renormalized the constant so that the $m_s \rightarrow \infty$ limit recovers GR.} of~\cite{Hou:2017cjy}, we find 
\begin{equation}\label{eq:SEPEq}
    \Delta = \frac{(q_1 - q_2)q_3}{G_{4(0,0)}(G_{4(0,0)}+1) \zeta}\left(1+\frac{r}{\lambdabar}\right)e^{- r/\lambdabar} \,,
\end{equation}
where the mass of the scalar field inside the Compton length $\lambdabar$ is defined as 
\begin{equation}
m_s^2 \equiv - \frac{G_{2(2,0)}}{\zeta}\,,
\end{equation}
with 
\begin{equation}
G_{i(m,n)} \equiv \frac{\partial^{m+n}G_i (\phi,X)}{\partial \phi^m \, \partial X^n}\bigg|_{\phi=\phi_0,X=0}\,.
\end{equation}

\begin{figure}
    \centering
    \vspace{2mm}
    \includegraphics[width=0.6\linewidth]{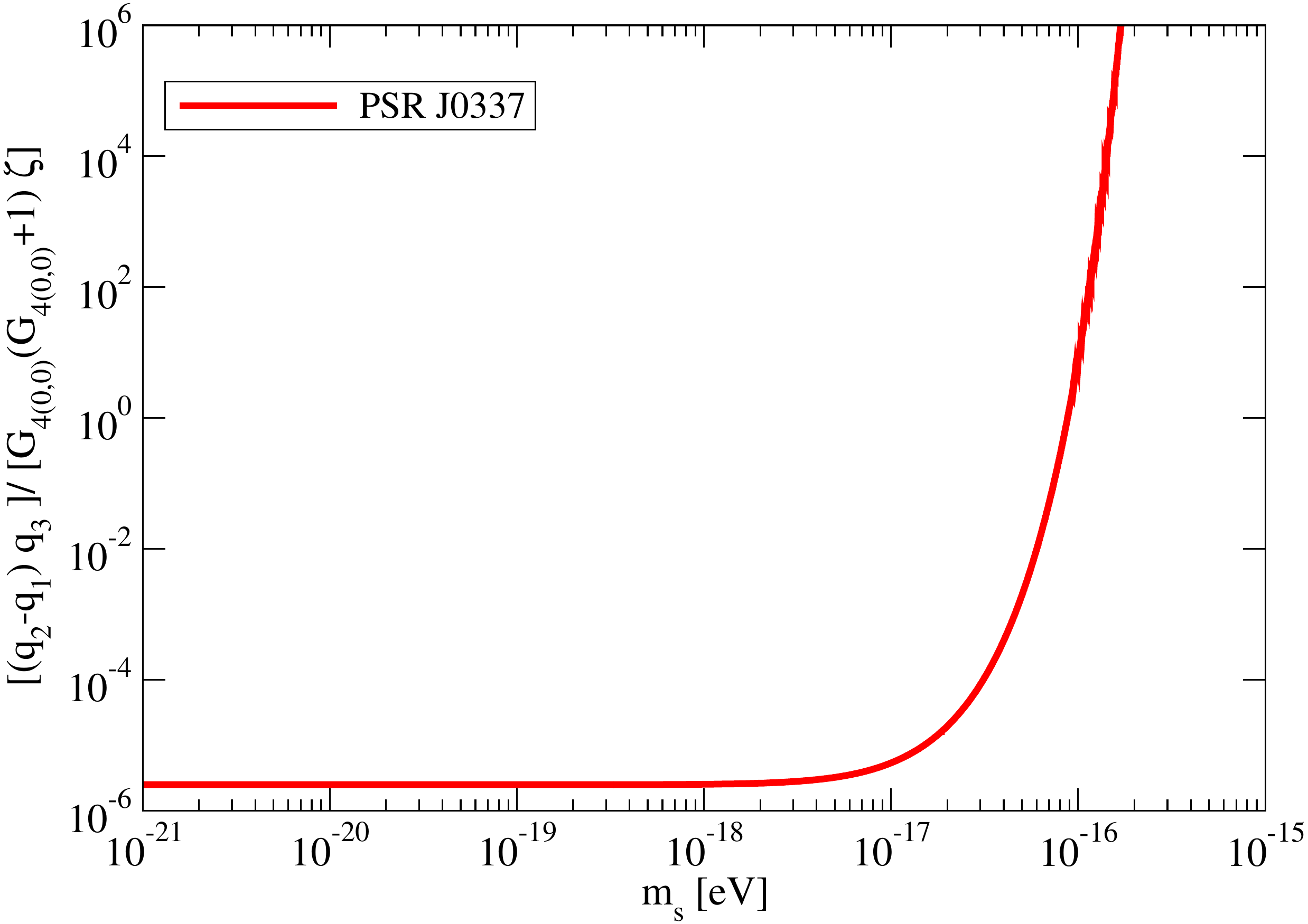}
    \caption{Lower bound on $(q_2-q_1)q_3/(G_{4(0,0)}(G_{4(0,0)}+1) \zeta)$ as a function of the mass of the scalar field from PSR J0337. }
    \label{fig:horn}
\end{figure}

The dimensionless scalar charge $q_i$ of body $i$ and parameter $\zeta$ in Eq.~\eqref{eq:SEPEq} are defined by
\begin{align} \label{eq:SEPCts}
    q_i &= G_{4(1,0)} -\frac{2 s_i}{\phi_0}G_{4(0,0)} \, , \\
    \zeta &= G_{2(0,1)}-2G_{3(1,0)}+3\frac{G_{4(1,0)}^2}{G_{4(0,0)}}\, . 
\end{align}
Figure~\ref{fig:horn} presents the bound on $(q_2-q_1)q_3/(G_{4(0,0)}(G_{4(0,0)}+1) \zeta)$ against $m_s$ from the fifth force measurement of PSR J0337. 

We now comment on how one can derive the massive Brans-Dicke result from the above Horndeski one. Horndeski theory reduces to massive Brans-Dicke theory under the following choice of the arbitrary functions~\cite{McManus:2016kxu}:
\begin{equation}
    G_2 = \frac{2 \omega_\BD}{\phi} X - U(\phi) ; \quad G_4 = \phi;\quad G_3=G_5 = 0 \, ,
\end{equation}
in conjunction with the definition for $\phi_0$ in massive Brans-Dicke given by ~\cite{Berti:2012bp}
\begin{equation}
    \phi_{0}=\frac{4+2 \omega_{\BD}}{3+2 \omega_{\BD}} \, .
\end{equation}
Substituting these into Eq.~\eqref{eq:SEPEq} and using Eq.~\eqref{eq:SEPCts}, we arrive at
\begin{equation}
	\Delta = \frac{2}{3+\omega_\BD}\left(1+ \frac{r}{\lambdabar_s} \right) (s_2-s_1)(1-2s_3) e^{- r/\lambdabar_s} \, .
\end{equation}
Thus, we have successfully recovered the expression for massive Brans-Dicke theory from that in Horndeski gravity. In the end then, this constraint on Horndeski gravity is powerful because it can be mapped onto a large class of scalar-tensor theories so that many can be tested.

\subsection{Axions}
In theories with axions, compact objects may acquire scalar charges that give rise to a fifth force~\cite{Hook:2017psm,Huang:2018pbu}. 
The axion field is sourced by large densities such as those inside stellar objects, including pulsars and WDs in PSR J0337. 
The scalar (or axion) charges are found by the equation in Table~\ref{tab:Theories} from~\cite{Poddar:2019zoe,Hook:2017psm,Huang:2018pbu} where $R_i$ is the radius and $\rho_i$ is the density of the $i$th body, while $m_s$ is the axion mass and $f_a$ is the axion decay constant. However, note that the axion field can only be sourced at a critical density such that $\rho_i  > m_s^2 f_a^2/\hbar^3$ and $\rho_i  >f_a^2/\hbar R_i^2 $ \cite{Hook:2017psm}. 
Because axion charge decreases with larger compactness, the WDs will dominate strength of the axion field compared to neutron stars in PSR J0337 (because $1/\ln(1-2 m/R)$ goes to zero as $m/R$ approaches $1/2$.). 
Using the values of the axion scalar charges, we find the expression for the SEP violation parameter $\Delta$ shown in Table~\ref{tab:Theories}.

\begin{figure}[thb]
    \centering
    \includegraphics[width=0.5\linewidth]{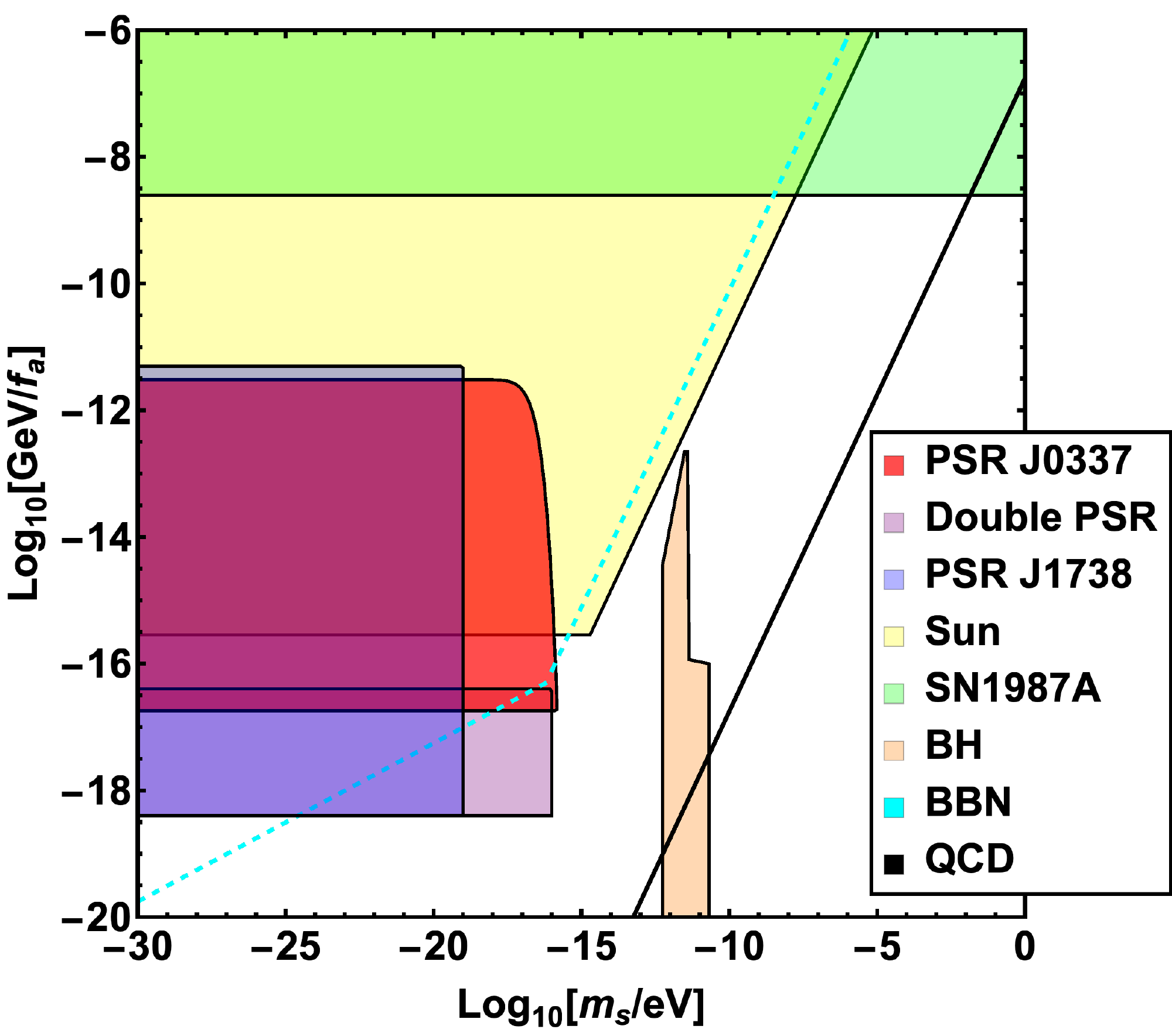}
    \caption{
   Excluded regions of the axion parameter space from various observations. We show the constraints due to the absence of a fifth force in PSR J0337 (red) and the orbital decay measurement of the double-pulsar binary \cite{Hook:2017psm} and PSR J1738 \cite{Poddar:2019zoe} (magenta and blue). We also show the constraints from measurements of the sun (yellow), supernova SN1987A (green), and black hole (BH) spins (orange)~\cite{Hook:2017psm,Arvanitaki:2014wva,Arvanitaki:2010sy}. If the axion is the dark matter source, the region above the cyan line is ruled out due to Big Bang Nucleosynthesis (BBN) constraints~\cite{Blum:2014vsa}. The QCD axion with parameters finely tuned to solve the strong CP problem must lie on the black line~\cite{Huang:2018pbu}. 
}
    \label{fig:axion}
\end{figure}

Figure~\ref{fig:axion} presents the constraints in the parameter space for the axion. The shaded regions are those which are excluded by observations. Notice that PSR J0337 can exclude a large region of the parameter space. In particular, we can now close the previously allowed gap between the bounds from binary pulsar and solar observations. Notice also that the shape of the excluded region for axions from PSR J0337 is similar to that for massive Brans-Dicke theory in Fig.~\ref{fig:MBDOmegaBound}, except there is also a minimum line at $1/f_a =  10^{-17.5}$GeV$^{-1}$ due to the density becoming lower than one of the critical ones.
Notice that the range of the axion mass that can be probed from pulsar observations is less than $\sim 10^{-16} \text{eV} $. Thus, the pulsar observations probe a different regime than those from e.g.~the axion dark matter detection experiment ABRACADABRA examining $10^{-12} \text{eV} \lesssim m_s \lesssim 10^{-6} \text{eV}$~\cite{Ouellet:2019tlz,Ouellet:2018beu}. Thus, these two methods for constraining the axion parameter space are complementary.

\subsection{Light Force Mediators of Dark Matter}
Finally, let us consider fifth forces due to light force mediators of dark matter. Dark matter can become gravitationally bound inside neutron stars or WDs~\cite{Goldman:1989nd,Kouvaris:2010vv}. Moreover, a massive light force mediator could cause a fifth force between bound dark matter in the stars~\cite{Croon:2017zcu,Ellis:2017jgp,Kopp:2018jom}.

We test for a light force mediator between two stars generically following Ref.~\cite{Alexander:2018qzg}. First, we assume a scalar field potential given by Eq.~\eqref{eq:YukawaPotential} and define $\alpha \equiv B q_iq_j$ as the interaction strength due to the light force mediator~\cite{Alexander:2018qzg}. 
The value of $\alpha$ depends not only on the dark matter model but also on the type of objects involved. 
We show the expression for the fifth force parameter $\Delta$ arising from this potential in Table~\ref{tab:Theories}. 
We simplify this by assuming that more dark matter will be bound in a neutron star than in a WD. This assumption suggests that the interaction between the pulsar and WD will dominate that of the two WDs: $\alpha_{\PSRWD}\gg \alpha_{\WDWD}$. This assumption is a consequence of the NS having a higher dark matter capture rate. The dark matter capture rate is proportional to both baryon density and escape velocity squared, so the bound dark matter in a NS will dominate that of the WD \cite{McDermott:2011jp}. From this, we can constrain the value of $\alpha_{\PSRWD}$ with the measurement of PSR J0337. 

\begin{figure}[thb]
    \centering
    \includegraphics[width=0.6\linewidth]{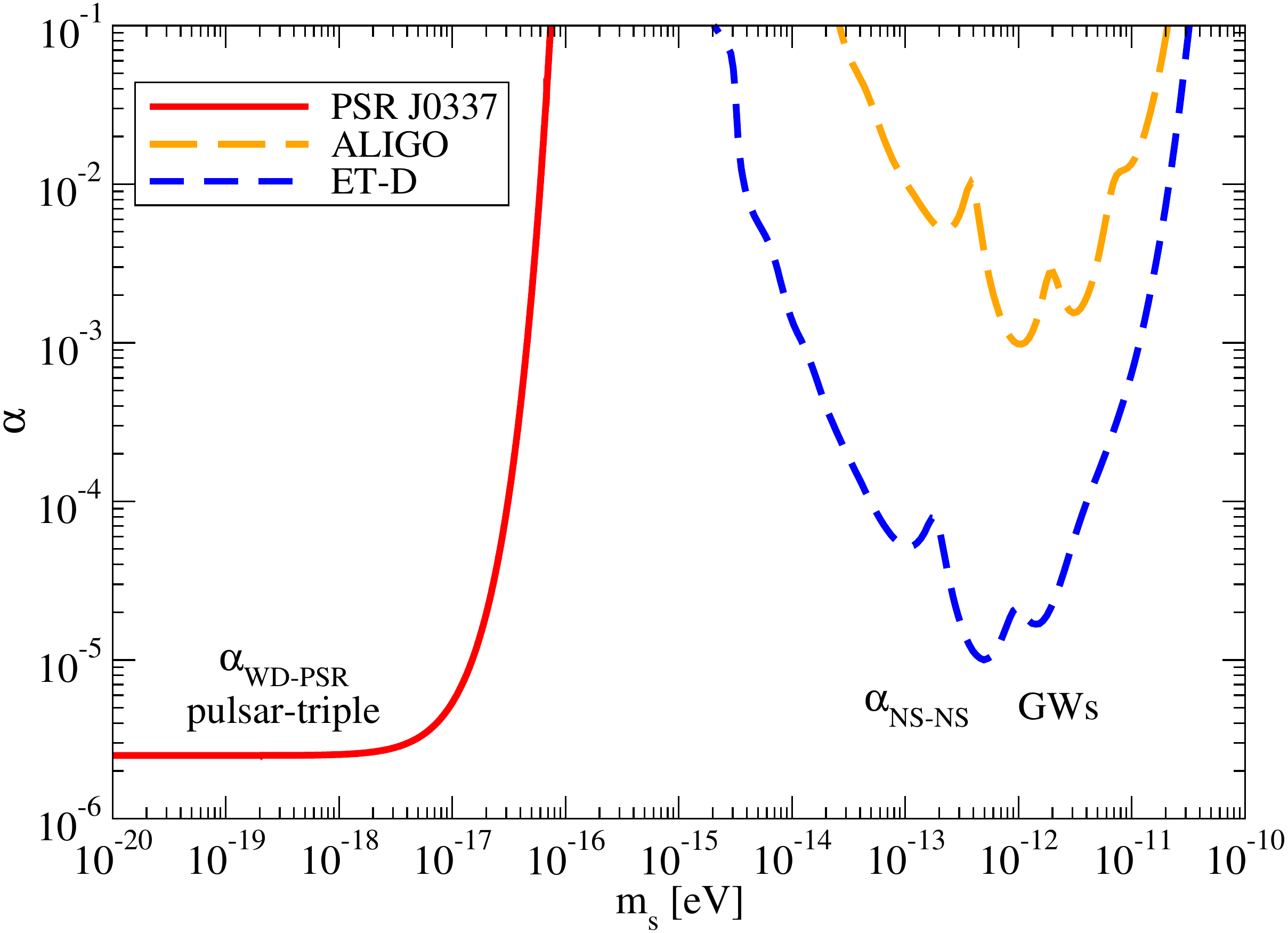}
    \caption{
    The upper bound of the dark matter Yukawa coupling for binaries as a function of scalar field mass from PSR J0337 and future gravitational-wave detections. On the left, we show the constraints for the coupling of a WD-pulsar binary from fifth force measurements of PSR J0337 (solid red). On the right, we show the upper bound on the coupling of a NS-NS binary with future gravitational wave observations with advanced LIGO (dashed orange) and Einstein Telescope (dashed blue)~\cite{Alexander:2018qzg}. The GW bounds come with a representative system with masses of $(2.0 M_\odot, 1.4 M_\odot)$ at distance of 100 Mpc. The full parameters used in the Fisher analysis is in Tab.~I of Ref.~\cite{Alexander:2018qzg}. Notice that the bounds from two different observations are complementary to each other.
    }
    \label{fig:dmconstraint}
\end{figure}

Figure~\ref{fig:dmconstraint} presents the lower bound on $\alpha_{\PSRWD}$ as a function of the mass of the light force mediator. Notice that the shape of the curve corresponds to flipping the one in Fig.~\ref{fig:MBDOmegaBound} upside down. This is because the former is showing the lower bound while the latter is showing the upper bound. We also present constraints on $\alpha_\NSNS$ from future gravitational-wave detections for comparison~\cite{Alexander:2018qzg}. In terms of the magnitude, these pulsar bounds are comparable to those that will be obtained with future gravitational-wave detections~\cite{Alexander:2018qzg}, though they are complementary as the mass range being probed is different due to the different size of the binary systems.

\section{Conclusion and Discussion}
In this paper, we studied how well one can probe massive scalar fields with the fifth force measurement in the pulsar triple system J0337. We have developed a formalism to describe the fifth force effect for massive scalar fields due to an additional Yukawa potential in a generic way. Applying this to various example theories, we found that we can place the current strongest bound on massive Brans-Dicke theory. We have also found that we can restrict new regions of the axion parameter space. Furthermore, we have placed new constraints on the force between a neutron star and WD due to interactions from bound dark matter. Lastly, our expression for SEP bounds in Horndeski theory is generic and can be used to further test other scalar-theories.

The future is bright for the discovery of more hierarchical systems to further constrain a fifth force and SEP violation. So far, estimates suggest that only about 5\% of the pulsars in the Milky Way have been discovered~\cite{Smits:2008cf}. New radio telescopes such as FAST~\cite{Nan:2011um} and SKA~\cite{Carilli:2004nx} will soon come online and will vastly improve sensitivity. The results presented here show that a fifth force or SEP violation is and will continue to be a powerful way of testing GR. Our generic formalism for massive scalar fields can easily be applied to such future detections.

\emph{Acknowledgments}
We thank Emanuele Berti, Scott Ransom, Lijing Shao and Clifford Will for useful discussions and valuable comments. 
We also thank Junwu Huang for clarifying axion charges.
K.Y. acknowledges support from NSF Award PHY-1806776. 
K.Y. would like to also acknowledge networking support by the COST Action GWverse CA16104. 

\bibliographystyle{iopart-num}
\bibliography{bibliography.bib}

\end{document}